\documentclass[11pt,a4paper]{article}
\input{epsf}

\catcode`\@=11

\def\thebibliography#1{\section*{REFERENCES}\list{\arabic{enumi}.}
  {\settowidth\labelwidth{#1.}\leftmargin=1.67em
   \labelsep\leftmargin \advance\labelsep-\labelwidth
   \itemsep\z@ \parsep\z@
   \usecounter{enumi}}\def\makelabel##1{\rlap{##1}\hss}%
   \def\newblock{\hskip 0.11em plus 0.33em minus -0.07em}
   \sloppy \clubpenalty=4000 \widowpenalty=4000 \sfcode`\.=1000\relax}

\def\@cite#1#2{${}^{{#1\if@tempswa , #2\fi})}$}

\newcount\@tempcntc
\def\@citex[#1]#2{\if@filesw\immediate\write\@auxout{\string\citation{#2}}\fi
  \@tempcnta\z@\@tempcntb\m@ne\def\@citea{}\@cite{%
	\@ordonner{#2}%
	\@for\@citeb:=#2\do%
    {\@ifundefined{b@\@citeb}%
	{\@citeo\@tempcntb\m@ne\@citea%
        	\def\@citea{,\penalty\@m\ }{\bf ?}\@warning%
       		{Citation `\@citeb' on page \thepage \space undefined}}%
    	{\setbox\z@\hbox{\global\@tempcntc0\csname b@\@citeb\endcsname\relax}
     \ifnum\@tempcntc=\z@ \@citeo\@tempcntb\m@ne%
       \@citea\def\@citea{,\penalty\@m}%
       \hbox{\csname b@\@citeb\endcsname}%
     \else%
      \advance\@tempcntb\@ne%
      \ifnum\@tempcntb=\@tempcntc%
      \else\advance\@tempcntb\m@ne\@citeo%
      \@tempcnta\@tempcntc\@tempcntb\@tempcntc\fi\fi}}\@citeo}{#1}}%

\def\@citeo{\ifnum\@tempcnta>\@tempcntb\else\@citea
  \def\@citea{,\penalty\@m}%
  \ifnum\@tempcnta=\@tempcntb\the\@tempcnta\else
   {\advance\@tempcnta\@ne\ifnum\@tempcnta=\@tempcntb \else
\def\@citea{-}\fi
    \advance\@tempcnta\m@ne\the\@tempcnta\@citea\the\@tempcntb}\fi\fi}

\def\@toto{}
\newif\if@ordre 
\newcount\c@current
\newcount\c@last

\def\@ordonner#1{\global\c@last\m@ne%
		\global\@ordretrue%
		\@for\@toto:=#1\do%
			{\@ifundefined{b@\@toto}%
			{}%
			{\c@current\csname b@\@toto\endcsname\relax%
			\ifnum\the\c@current<\the\c@last\relax%
				{\global\@ordrefalse}\fi%
			\global\c@last\the\c@current%
			}%
			}%
		\if@ordre{}\else{\typeout{}%
			\typeout{Warning: the references are not %
			 in increasing order\on@line:}%
			\@for\@toto:=#1\do%
			{\@ifundefined{b@\@toto}%
			{}%
			\typeout{\@toto:\space \@nameuse{b@\@toto}}%
			}\typeout{}}\fi%
		}%

\catcode`\@=12

\newcommand{\slL}{\raise.15ex\hbox{$/$}\kern-.53em\hbox{$L$}}
\newcommand{\slP}{\raise.15ex\hbox{$/$}\kern-.53em\hbox{$P$}}
\newcommand{\slR}{\raise.15ex\hbox{$/$}\kern-.53em\hbox{$R$}}
\newcommand{\slQ}{\raise.15ex\hbox{$/$}\kern-.53em\hbox{$Q$}}
\newcommand{\slK}{\raise.15ex\hbox{$/$}\kern-.53em\hbox{$K$}}

\newcommand{\be}{\begin{equation}}
\newcommand{\ee}{\end{equation}}     
\newcommand{\bea}{\begin{eqnarray}}
\newcommand{\ena}{\end{eqnarray}}

\def\build#1\over#2{\mathrel{\mathop{\kern 0pt#2}\limits_{#1}}}

\font\tenimbf=cmmib10 at 12pt
\font\sevenimbf=cmmib10 at 7pt
\font\fiveimbf=cmmib10 at 5pt
\newfam\imbf
\textfont\imbf=\tenimbf
\scriptfont\imbf=\sevenimbf
\scriptscriptfont\imbf=\fiveimbf
\def\imb{\fam\imbf\tenimbf}

\begin{document}

\vglue 4cm
\centerline{\Large\bf{PHOTONS, DILEPTONS}}
\def\thefootnote{\fnsymbol{footnote}}
\centerline{\Large\bf{AND HARD THERMAL LOOPS \footnote{Work done in 
collaboration with P.~Aurenche, R.~Kobes and
E.~Petitgirard.} 
\footnote{Talk
given at the XXXIInd Rencontres de Moriond, Les Arcs, 22-29 March 1997.}
}}
\vskip 1cm

\begin{center}
{\bf Fran\c cois Gelis}\\
Laboratoire de Physique Th\'eorique ENSLAPP,\\
URA 1436 du CNRS associ\'ee \`a l'Ecole Normale Sup\'erieure de Lyon et \`a l'Universit\'e de
Savoie,\\
B.P. 110, F-74941 Annecy-le-Vieux Cedex, France
\end{center}

\vfill
\centerline{\bf Abstract}
The production rate of soft real photons or soft lepton pairs by a hot QCD
plasma is dominated by strong collinear divergences. As a consequence, it
appears that the effective theory based on the resummation of hard thermal
loops fails to handle properly these light-cone sensitive processes since some
formally higher order diagrams are in reality the dominant ones.
\vskip 1cm
\centerline{\hfill ENSLAPP-A-646/97}
\centerline{\hfill hep-ph/9705360}
\vfill
\eject

\section{Introduction}
\subsection{Plasma observables}
There are several quantities that could be interesting probes for a
quark-gluon
plasma. Let us briefly describe a few such possibilities\footnote{In this 
theoretical description, we present only ``ideal'' experiments, in the sense 
that we  assume having at our disposal a sufficient amount of plasma at
thermal  equilibrium. Unfortunately, they will remain thought experiments for a
long time, and it is therefore important to develop also theoretical tools to
study the  real world experiments. This requires to deal with
out-of-equilibrium plasmas, a subject beyond the scope of the present talk.},
recalling the  theoretical tools used for the calculation of each of them. The
first possibility would be to measure the thermodynamic properties of such a
plasma,  by measuring for instance its pressure as a function of temperature
and volume. From a theoretical point of view, this quantity requires the
calculation of the free energy, which has been achieved up to the order $g^5$,
where $g$ is the  strong coupling constant, in thermal QCD. Another interesting
experiment would be consist in studying the scattering of an external hard
particle by a plasma. Since this situation involves the interaction of the
plasma with some non thermalized object, it cannot be dealt with by the
standard methods of thermal field theory. So far, it has been studied by semi
classical methods. The last observable I have selected is the production rate
of some weakly interacting particle, like photons or lepton pairs. This quantity 
has been
calculated both  by semi-classical methods \cite{cgr} and by thermal field
theory \cite{bpy,bps,agkp}. The
main part of this talk is devoted to the latter, while a comparison with the
results of semi classical methods is presented briefly at the end.

\subsection{Thermal field theory tools for the photon rate}
The real photon production rate, per unit time and per unit volume of the 
plasma, is related simply to the thermal photon 
polarization 
tensor by:
\begin{displaymath}
q_0 {{dN}\over{d^4x d^3{\imb q}}}= -{1\over{(2\pi)^3}} n_{_{B}}(q_0) {\rm Im}
\,
\Pi^\mu{}_\mu(q_0,{\imb q})\;,
\end{displaymath}
where $n_{_{B}}(q_0)\equiv 1/(\exp(q_0/T)-1)$ is the Bose-Einstein function. The
production rate of lepton pairs (or, equivalently of virtual photons), is also
related to  ${\rm Im}
\,
\Pi^\mu{}_\mu(q_0,{\imb q})$, with a somewhat different phase space.
It is worth saying that this formula is valid to all orders in the
strong coupling constant, but only to first order in $\alpha$, since 
effects like the re-interactions of the emitted photons in the medium
are neglected. This is
justified by the smallness of the electromagnetic coupling constant.

\subsection{Infrared problems and hard thermal loops summation}
Nevertheless, the game consisting in the calculation of  ${\rm
Im}\,\Pi^\mu{}_\mu$ in thermal QCD is not as simple as it seems,
because thermal field  theories are plagued by infrared singularities. One may
easily understand why these singularities can be stronger than at $T=0$, by
noticing that a Bose-Einstein statistical factor can be very large for soft
energies: $n_{_{B}}(l_0)\sim T/l_0 \gg 1$ for $l_0\ll T$. In the following, it is
useful to distinguish between two energy scales: the hard  scale, of order $T$,
corresponds to the typical energy of partons in the plasma, and the soft
scale, of order $gT$, which is the typical scale of the quanta exchanged by
interacting partons\cite{bp}. Noticing that certain loops carrying hard momentum can be
as large as  their bare counterparts if all their external legs are soft, and
need therefore to be taken into account already to calculate consistently the
first order of perturbation theory,  Braaten and Pisarski set up an effective
theory resumming  these ``Hard Thermal Loops" (HTL in the following), while
preserving the gauge symmetry of the bare theory \cite{bp}. Physically, this resummation
is closely related to the Debye screening in a plasma: through the resummation
of HTLs, the gauge bosons can acquire a thermal mass $m_{\rm g}\sim gT$, which
is nothing but the inverse of the range of the screened interaction.
 
\subsection{Further problems}
Despite its nice features, the effective theory one gets after the resummation
of HTLs has still a few problems that are not solved by this resummation. The
first one is related to the absence of thermal mass for the static transverse
gauge bosons. In QED, this is known to be true to all orders, and is physically
related to the fact that static magnetic fields are not screened in a plasma.
In QCD, the status of this ``magnetic mass" is not so clear since the self
interaction of gauge bosons can generate such a mass. Nevertheless, if such a
mass exists, it is expected to be at most of order $g^2T$ and beyond the
abilities of perturbative calculations.

Another problem that is not solved by the HTL resummation is related to
collinear singularities. Technically, this is due to the fact that the
building blocks of the HTLs are bare massless propagators. As a consequence,
the HTLs, which are the building blocks of the effective theory, become
singular when some of their external legs are put on the light--cone, due to
angular integrals like $\int {d\cos\theta}/{E(1-\cos\theta)}$. A solution to
overcome these singularities consists in taking into account an asymptotic
thermal mass $m_{_{F}}$ 
for the hard fermions that run inside the HTL \cite{fr}. Na\"\i vely, the
previous integral now becomes $\int {d\cos\theta}/{(E-p\cos\theta)}$,  with
$E=\surd(p^2+m_{_{F}}^2)$. 
The important result lies in the fact that such a procedure
preserves gauge invariance while regularizing the collinear divergences.

\section{Photon production rate at 1 loop in the effective theory}
\subsection{Results of previous calculations}
At one loop in the effective theory, there are $3$ diagrams contributing to the
photon production rate. For each of them, we give some of the relevant 
physical amplitudes, as well as the result of previous calculations
for real photons \cite{bps} (for dileptons, see \cite{bpy}):
\vskip 2mm

\hbox to 
\textwidth{\raise 2cm\vbox{\hbox{${\imb [a]:\;}$}}\vbox{\epsfbox{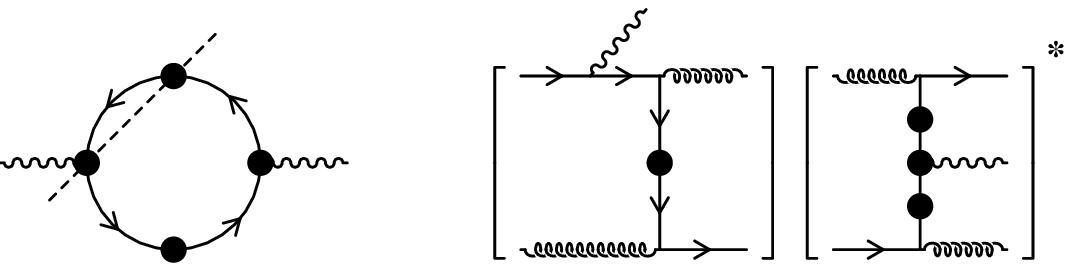}}
\raise 1cm\vbox{\hbox{$\qquad{\rm Im}\,\Pi^\mu{}_\mu(Q)_{|{\imb [a]}}
\sim e^2 g^4 {{T^3}
\over{q_0}}$}}\hfill}
\vskip 5mm
\hbox to 
\textwidth{\raise 2cm\vbox{\hbox{${\imb [b]:\;}$}}\vbox{\epsfbox{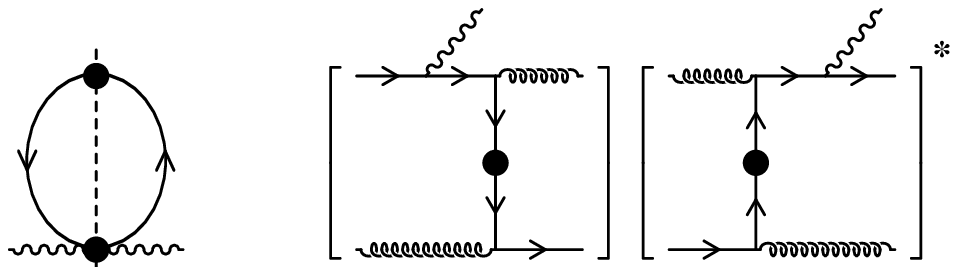}}
\raise 1cm\vbox{\hbox{$\qquad{\rm Im}\,\Pi^\mu{}_\mu(Q)_{|{\imb [b]}}
\approx 0$}}\hfill}
\vskip 5mm
\hbox to 
\textwidth{\raise 2cm\vbox{\hbox{${\imb [c]:\;}$}}\vbox{\epsfbox{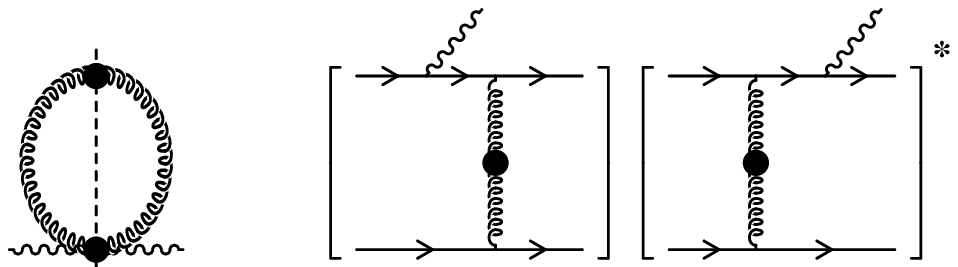}}
\raise 1cm\vbox{\hbox{$\qquad{\rm Im}\,\Pi^\mu{}_\mu(Q)_{|{\imb [c]}}
\approx 0$}}\hfill}

A few comments are useful concerning these results. First of all, the only non 
vanishing term  ${\imb [a]}$ corresponds to a very complicated physical
amplitude, which is not intuitive at all. Secondly, the result found for ${\imb
[a]}$ is  smaller than what could be expected for soft photons by simple power
counting.  This fact seems to be due to a cancelation particular to the trace
$\cdots^\mu{}_\mu$, since it does not occur in ${\rm Im}\,\Pi^{00}$ for
instance. Therefore, one should expect that this result is not complete since
the approximations made in the calculation were designed to deal
with dominant terms, whereas the first
non vanishing order for  ${\rm Im}\,\Pi^\mu{}_\mu$ is subdominant because of
this cancelation.

As a consequence, one should perform the calculation of the same diagrams {\sl 
beyond the HTL approximation} \cite{agkp}, including ${\imb [b]}$ and ${\imb [c]}$, 
since 
the vanishing result obtained so far is only a consequence of the HTL 
approximation associated to the trace $\cdots^\mu{}_\mu$, that does not tell us 
anything about subdominant terms. In reality, by noticing that ${\imb [b]}$ 
involves a Fermi-Dirac weight evaluated at a soft energy whereas ${\imb [c]}$
involves a Bose-Einstein weight at the same energy, one can avoid the
calculation of  ${\imb [b]}$, which is negligeable in front of ${\imb [c]}$.
Moreover, it turns out that the result for ${\imb [c]}$ is much greater than
the partial result already obtained for ${\imb [a]}$, which means that only 
${\imb [c]}$ is needed.

\subsection{Diagram {\bf (c)} beyond the HTL approximation}
In order to perform the calculation, we come back to the diagrammatic expansion
of  the effective $\gamma-\gamma-g-g$ vertex, which amounts to the following two
topologies: \vskip 3mm \hbox to \textwidth{\vbox{\epsfbox{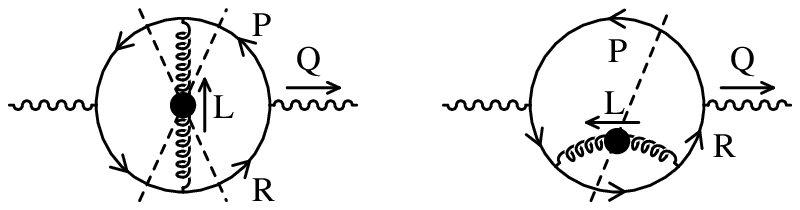}}\hfill} The
first step is now to perform the Dirac's algebra associated to the fermion loop.
Na\"\i vely, for a loop with four hard fermions, the Dirac's traces should be of
order  $T^4$. Nevertheless, due to the $\cdots{}^\mu{}_\mu$ cancelation, these
traces are suppressed and their order of magnitude is in reality $g^2T^4$. More
precisely,  these traces are ${\imb r}^2L^2$ in the  case of the vertex
correction topology, and  ${\imb r}^2Q^2$ for the self energy correction.
Therefore, if the produced photon is close enough to the light-cone, we can
neglect the self-energy diagram with respect to the vertex correction.

Another feature of the vertex diagram lies in the strong collinear singularities
that can potentially appear when the photon is emitted collinearly to the quark.
Indeed, the imaginary  part of this diagram contains the following factors,
relevant for these considerations:
${{\delta(P^2)\delta((R+L)^2)}/{(P+L)^2(P+Q)^2}}$, where we omitted for the sake
of simplicity the fermion thermal mass. We easily see that the dangerous
denominators, both of them leading to a simple pole in $\cos\theta$ where
$\theta$ is the angle between the quark and the photon, can vanish almost
simultaneouly since ${\imb p}$ and ${\imb r}+{\imb l}$ are collinear to ${\imb
q}$ almost at the same time (${\imb p}$, ${\imb r}$ are hard, whereas ${\imb q}$
and ${\imb l}$ are soft). As a consequence, these factors behave very much like a
 double pole, which means that the regulator (a thermal mass $m_{_{F}}\sim
gT$) will appear as $T^2/m_{_{F}}^2$ instead of inside a logarithm as it would be
for two separate simple poles, and this fact can change the order of magnitude of
the whole result. More precisely, taking the fermion thermal mass into account,
and being very rough with algebra, such a double pole will give\footnote{This is
the basic structure of the angular integral when the emitted photon is massless
($Q^2=0$). In the case $Q^2>0$, a more precise study of the kinematics shows that
it is sufficient to replace in this integral the fermion thermal mass by an
effective mass taking into account the regularizing effect of a positive $Q^2$:
$m^2_{\rm eff}=m_{_{F}}^2+Q^2{\imb r}^2/q_0^2$.}: $\int\nolimits_{-1}^{1}
{{d\cos\theta}/{\left[1-\cos\theta+m_{_{F}}^2/2{\imb r}^2\right]^2}}\sim {{{\imb
r}^2}/{m_{_{F}}^2}}\sim {1/{g^2}}\gg 1$, whereas such a dimensionless angular
integral would have been of order $1$ in the absence of collinear singularities,
and at most of order $\ln(r^2/m_{_{F}}^2)$ in the case of separate simple poles.
Let us now summarize the features of the result we get for this diagram:
   
   (i) ${\rm Im}\,\Pi^\mu{}_\mu(Q)_{|{\imb [c]}}\approx -\sum\limits_{_{T,L}}^{}
   e^2  g^2 J_{_{T,L}} {{T^3}\over{q_0}}\sim e^2 g^2 {{T^3}\over{q_0}}$, where
   $J_{_{T,L}}$ is a numerical factor  depending on the two dimensionless ratios
   $m_{\rm g}^2/m_{_{F}}^2$ and $Q^2T^2/q_0^2 m_{\rm g}^2$, where $m_{\rm g}$ is
   the gluon thermal mass. Moreover, the transverse gluon contribution is of the
   same importance as the longitudinal one. The functions $J_{_{T,L}}$ are
   plotted on the following left figure, for $m_{\rm g}^2/m_{_{F}}^2=1.5$ and
   $m_{\rm g}^2/T^2=0.1$. We see that the result is considerably enhanced in the
   region of very small photon invariant mass.

   (ii) The result is free of any infrared divergence, even for the transverse
   gluon exchange. As a consequence, there is no need for a magnetic mass. In
   fact, including such a mass has no effect on the result provided that $m_{\rm
   mag}\ll m_{_{F}}$. The effect of $m_{\rm mag}$ has been evaluated numerically
   on the following right figure, at $Q^2=0$, for $m_{\rm g}^2/m_{_{F}}^2=1$
   (solid line) and $m_{\rm g}^2/m_{_{F}}^2=10$ (dotted line). \vskip 3mm \hbox
   to \textwidth{\vbox{\epsfbox{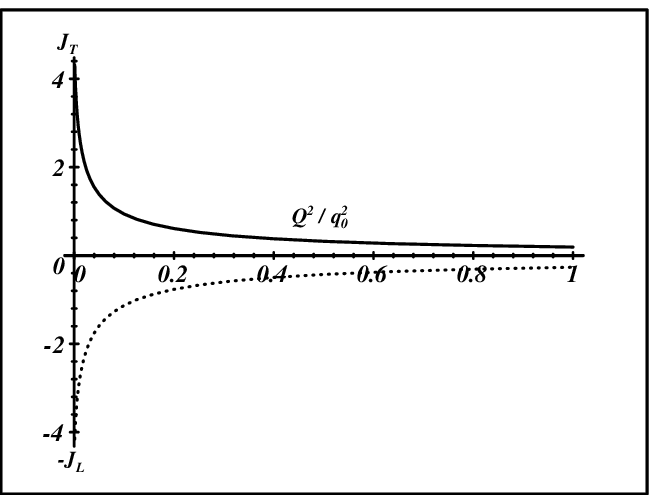}}\hfill\vbox{\epsfbox{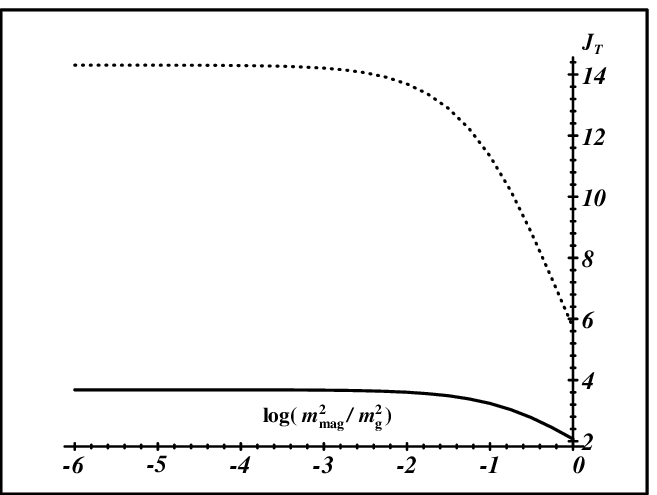}}\hfill}

   (iii) If $Q^2$ increases, the scattering becomes sensitive to the energy scale
   of $m_{\rm eff}$, intermediate between the hard and soft scales. This is a
   qualitative difference with the situations where the hard thermal loops
   framework holds.

\subsection{Comparison with semi-classical methods}
To be complete, it is instructive to compare this result, obtained in the
framework of thermal field theory, with the results obtained by means of
semi-classical approaches \cite{cgr}. To that purpose, we modified our
expressions in order to recover the standard semi-classical factorization:
\begin{eqnarray} {{dN}\over{d^4x}}&&\approx {{d^3{\imb q}}\over{(2\pi)^3 2q_0}}
\int {{d^4P_1}\over{(2\pi)^4}} {{d^4P_2}\over{(2\pi)^4}}
{{d^4P^\prime_1}\over{(2\pi)^4}} {{d^4P^\prime_2}\over{(2\pi)^4}} \;(2\pi)^4
\delta(P_1+P_2-P^\prime_1-P^\prime_2-Q)\nonumber\\ &&\times
(2\pi)^4\,n_{_{F}}(P_1^0)[1-n_{_{F}}(P^\prime_1{}^0)]n_{_{F}}(P_2^0)
[1-n_{_{F}}(P^\prime_2{}^0)]\, \delta(P_1^2-m_{_{F}}^2)
\delta(P^\prime_1{}^2-m_{_{F}}^2)\, \delta(P_2^2)\,
\delta(P^\prime_2{}^2)\nonumber\\ &&\times|{\cal
M}(P_1,P^\prime_1+Q;P_2,P^\prime_2)|^2\;\;\;e^2 \sum\limits_{{\rm pol. }
\epsilon}^{} \left({{P_1\cdot \epsilon}\over{P_1\cdot
Q}}-{{P^\prime_1\cdot\epsilon}\over{P^\prime_1\cdot Q}}\right)^2\nonumber
\end{eqnarray} where ${\cal M}$ is the amplitude corresponding to the same
scattering without photon emission, and the factor beginning by $e^2$ is the
square of the electromagnetic current that couples the photon to the quark. We
recognize in this formula the first term of the expansion performed in the
semi-classical methods, which seems to indicate that the terms beyond the HTL
approximation we have considered are actually relevant for the photo-emission
process by a hot plasma. The complete expansion in the semi-classical approach
gives the so called Landau Pomeranchuk Migdal suppression \cite{lpm}, but of
course, we cannot recover this effect here with only one scattering. On the other
hand, thermal QCD provides a much more rigorous framework to treat the transverse
gluons, and it is not obvious that the static scattering center approximation  at
the basis of the semi-classical approach is a good one, since it implies that the
transverse gluon exchange can be neglected which is not the case.

\section{Conclusions} The main conclusion of this work is that the HTL expansion
may breakdown in certain very specific circumstances, namely when effective
vertices have light-like external legs. This has dramatic consequences for the
calculation in thermal QCD of processes that are very sensitive to what happens
in
the vicinity
of the light--cone, which is precisely the case of photon production. In this
specific area, it appeared that the bremsstrahlung process is the dominant one,
which is in agreement with the results of semi-classical methods.

Future work on thermal photon production should certainly consider higher order
contributions, in order to determine whether the multiple scatterings are
important or not, thereby allowing a field theoretical approach towards the LPM
effect. On the other hand, from a more formal point of view, it would be
important to determine precisely what kind of reorganization of the perturbative
expansion would take care of the collinear enhancement encountered here. \vskip
3mm

\noindent{\bf Acknowledgments:}     It is a pleasure to thank the organizers of
such an enjoyable conference and P. Aurenche for his very helpful comments in
preparing this talk.


\begin{thebibliography}{99}

\bibitem{cgr} J.~Cleymans, V.~Goloviznin, K.~Redlich, 
                Phys.~Rev.~{\bf D47}, 989 (1993); Z.~Phys.~{\bf C59}, 495 (1993).
\bibitem{bpy} E.~Braaten, R.~~Pisarski, T.~C.~Yuan, 
                Phys.~Rev.~Lett.~{\bf 64}, 2242 (1990).
\bibitem{bps} R.~Baier, S.~Peign\'e, D.~Schiff, 
                Z.~Phys.~{\bf C62}, 337 (1994).
\bibitem{agkp} P.~Aurenche, F.~Gelis, R.~Kobes, E.~Petitgirard,
                Phys. Rev. D {\bf 54}, 5274 (1996); preprint hep-ph/9609256
		(to appear in Z. Phys. C {\bf 74}). 
\bibitem{bp} E.~Braaten, R.~Pisarski,
                Nucl.~Phys.~{\bf B337}, 569 (1990); {\bf B339}, 310 (1990).
\bibitem{fr} F.~Flechsig, A.~Rebhan, 
                Nucl.~Phys.~{\bf B464}, 279 (1996).
\bibitem{lpm} L.~Landau, I.~Pomeranchuk, Dokl.~Akad.~Nauk 
                {\bf 92}, 535 (1953); {\bf 92}, 735 (1953);
                A.~Migdal, Phys.~Rev.~{\bf 103}, 1811 (1956).




\end{thebibliography}
\end{document}